\documentclass{ws-ijmpcs}

\usepackage[super]{cite}
\usepackage{xcolor}
\usepackage[verbose,hypertexnames=false]{hyperref}
\hypersetup{colorlinks=false,allbordercolors=blue,pdfborderstyle={/S/U/W 1}}

\begin{document}

\markboth{S. Taheri Monfared}
{What can we learn from the Parton Branching method in QCD?}

%
\catchline{}{}{}{}{}
%

\title{What can we learn from the Parton Branching method in QCD?}

\author{Sara Taheri Monfared}

\address{Deutsches Elektronen-Synchrotron DESY, Germany\\
taheri@mail.desy.de}

\maketitle

\begin{history}
\end{history}

\begin{abstract}

This work reviews recent developments in the Parton Branching (PB) method, focusing on its application to Transverse Momentum Dependent (TMD) parton distributions and the implementation of TMD evolution equations in Monte Carlo generators. Key advancements include the inclusion of photon and heavy electroweak boson radiation in the evolution equations and their impact on collinear and TMD distributions. A detailed comparison of PB and Collins-Soper-Sterman formalisms highlights improvements in the accuracy of PB Sudakov form factors. The role of soft gluons, intrinsic transverse momentum, and the $z_M$ parameter in modeling non-perturbative effects is emphasized, with implications for inclusive distributions and Drell-Yan transverse momentum spectra. This review also addresses challenges in achieving consistency between forward and backward evolution.

\end{abstract}


\section{Parton Branching method}
The Parton Branching (PB) method provides evolution equations for Transverse Momentum Dependent (TMD) parton distribution functions (PDFs) and enables their application within TMD Monte Carlo (MC) generators. The TMD evolution equation is implemented and solved using MC techniques in the \textsc{uPDFevolv2} package \cite{Jung:2024uwc}, an extension of the earlier \textsc{uPDFevolv} framework \cite{Hautmann:2014uua}. The free parameters of the PB parton distributions at the initial evolution scale can be fitted to experimental data through the \textsc{xFitter} package \cite{Alekhin:2014irh,xFitter:2022zjb}. The TMD MC generator \textsc{CASCADE3} \cite{CASCADE:2021bxe,CASCADE:2010clj} incorporates these TMDs into event generation, providing a systematic procedure to match TMDs with fixed-order matrix elements \cite{BermudezMartinez:2018fsv,BermudezMartinez:2019anj}.
Here we review the latest developements of PB method to document the main lessens we learned from that. Here, we review the latest developments of the PB method to summarize the main lessons learned. 

\section{Heavy boson collinear and TMD distributions}

This section summarizes a solution of the extended DGLAP evolution equations, which include the photon and heavy electroweak bosons, based on the results of Refs.~\refcite{Jung:2024uwc,HEAVYBOSON}. 
Photon and heavy boson radiation from quarks occurs similarly to gluon radiation, with splitting functions identical to the QCD case but differing in color factors. Notably, photons lack a self-coupling process.

In the PB approach, the evolution equation is reformulated by replacing the plus-prescription with a Sudakov form factor, as discussed in detail in Refs.~\refcite{Hautmann:2017xtx,Hautmann:2017fcj}.
The evolution equation expressed in terms of the Sudakov form factor $\Delta^S_a(z_M, \mu^2)$ is given by:
\begin{equation}
  {x f}_a(x,\mu^2)  =  \Delta^S_a (  \mu^2  ) \  {x f}_a(x,\mu^2_0)  
+ \sum_b
\int^{\mu^2}_{\mu^2_0} 
{{d q^2 } 
\over q^2 } 
{
{\Delta^S_a (  \mu^2  )} 
 \over 
{\Delta^S_a( q^2
 ) }
}
\int_x^{z_{M}} {dz} \;
 \frac{\alpha_{eff}}{2\pi} \hat{P}_{ab} (z) \frac{x}{z}
\;{f}_b\left({\frac{x}{z}},
q^2\right)  \; ,
\end{equation}
where $\alpha_{eff}$ is defined in Refs.~\refcite{Jung:2024uwc,HEAVYBOSON}, and the indices $a$ and $b$ now also include the photon, Z boson, and W boson.
A key feature of this formulation is the limiting scale $z_{M}$ in the $z$-integral. 
To ensure consistency with the DGLAP equations for massless QCD partons and the photon, the integration limit is set to $z_{M} \to 1$. However, in numerical implementations, $z_{M}$ is generally chosen as $z_{M} = 1 - \epsilon$, where $\epsilon$ is a very small value. The value of $z_M$ is discussed in more detail in the next sections.

There are various methods to account for the masses of heavy bosons. Different treatments of the boson mass lead to distinct density distributions as a function of the evolution scale, which can have measurable effects. We found that the most effective approach is to incorporate the heavy boson mass as a suppression factor.

The collinear splitting functions for QCD partons are applied at next-to-leading order (NLO) accuracy. For the photon and heavy bosons, the leading-order (LO) splitting functions and couplings are used in the so-called "phenomenological" scheme. The initial distributions for quarks and gluons are fitted to deep inelastic scattering (DIS) precision data recorded at HERA. Notably, the inclusion of the photon and heavy bosons does not significantly change the fit parameters or the quality of the fit.
Additionally, the PB method enables the simultaneous determination of TMD densities for the photon and heavy bosons.

\begin{figure}[h!tb]
\begin{center} 
\includegraphics[width=0.49\textwidth,angle=0]{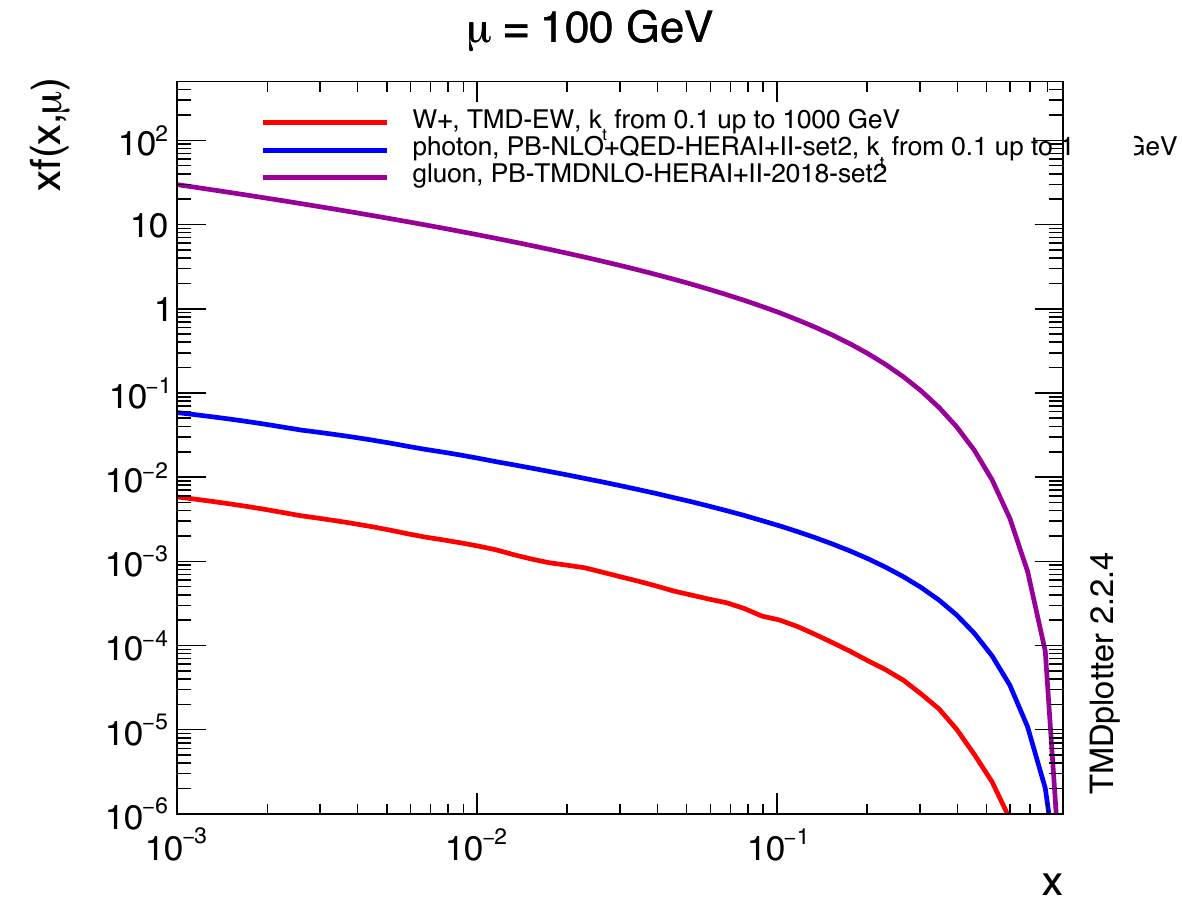}
\includegraphics[width=0.49\textwidth,angle=0]{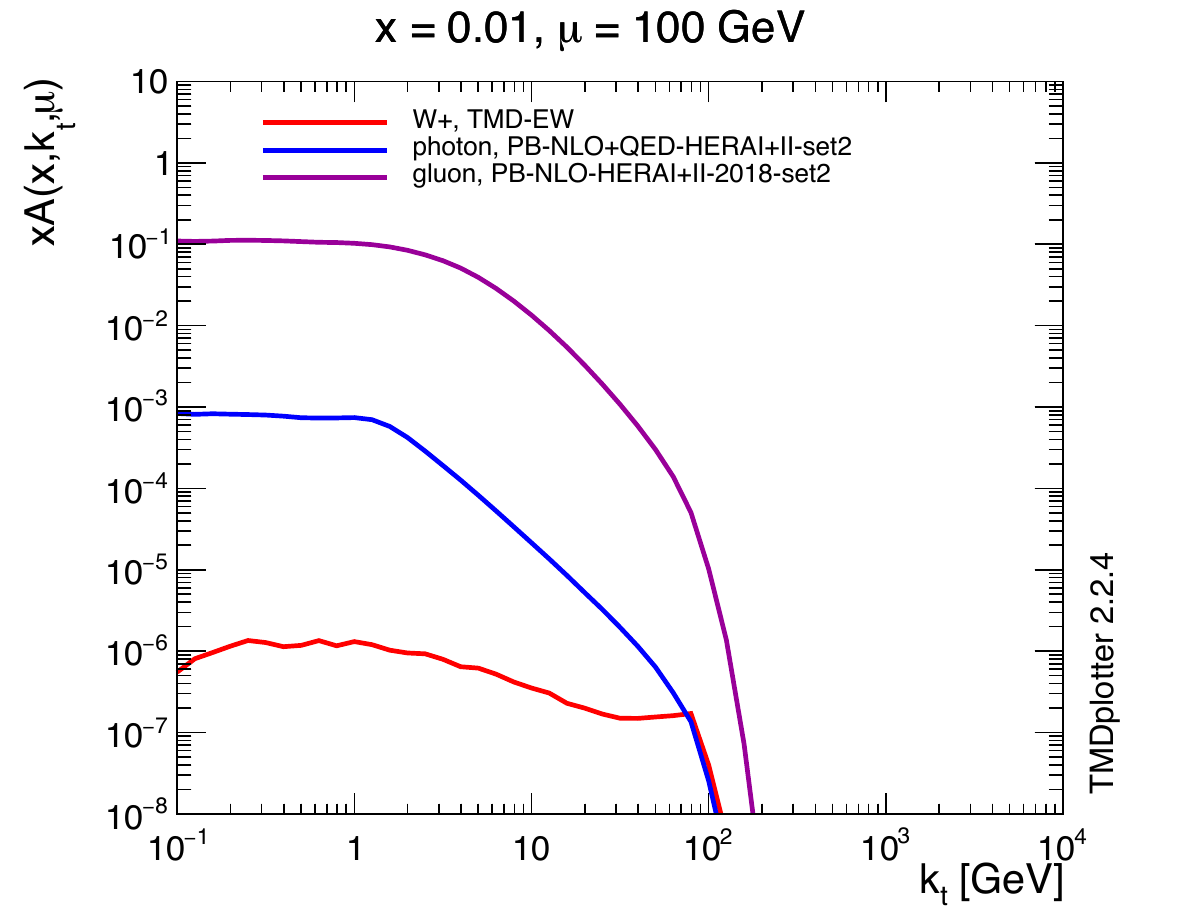}
\caption{\small The collinear and TMD vector boson densities  at $\mu= 100$ GeV as a function of $x$ and $k_T$ respectively (Plots taken from 
Ref. ~\cite{Jung:2024uwc}).
}
\label{EWFig2}
\end{center}
\end{figure}

In Fig.~\ref{EWFig2} (left plot), the heavy vector-boson densities are shown at large scales, as they vanish at scales below their masses. 
In the transverse momentum distribution, Fig.~\ref{EWFig2} (right plot), one can observe the similarity between the photon and $W$ densities at large $k_T$. However, significant differences are visible at smaller $k_T$, where the differences in collinear PDFs are well reflected.

\section{Logarithmic accuracy of the TMD PB method}
In this section, I summarized the results obtained in Ref. \refcite{CSS1,Lelek:2024kax}. 
The PB Sudakov form factor can be written in terms of virtual parts of the DGLAP splitting functions, using momentum sumrule.
\begin{equation} 
\Delta_a(\mu^2, \mu_0^2) \approx \exp\left( -\int_{\mu_0^2}^{\mu^2}\frac{\textrm{d}\mu^{\prime 2}}{\mu^{\prime 2}} \left( \int_0^{z_M} k_a(\alpha_s) \frac{1}{1-z} \textrm{d}z  - d_a(\alpha_s)\right)\right)\;.
\label{eq:VirtSud}
\end{equation}
Then they can be split into two parts by introducing an intermediate dynamical scale, $z_{\rm{dyn}}=1-q_0/{\mu^{\prime}}$, motivated by angular ordering defition
\begin{eqnarray}
\label{eq:divided_sud}
\Delta_a^{} ( \mu^2 , \mu^2_0 ) = && 
\exp \left(  -   
\int^{\mu^2}_{\mu^2_0} 
\frac{d \mu^{\prime 2} } 
{ \mu^{\prime 2} } \left[
 \int_0^{z_{\text{dyn}}(\mu')} dz 
  \frac{k_a(\alpha_s^{\rm{}})}{1-z} 
- d_a(\alpha_s^{\rm{}}) \right]\right)\nonumber \\
 & \times&  \exp \left(  -   
\int^{\mu^2}_{\mu^2_0} 
\frac{d \mu^{\prime 2} } 
{\mu^{\prime 2} } 
 \int_{z_{\text{dyn}}(\mu')}^{z_M} dz 
  \frac{k_a(\alpha_s^{\rm{}})}{1-z} 
\right)\;.\nonumber \\
 & =& \Delta_a ^ {(P)} ( \mu^2 , \mu^2_0 ) ~ \times ~ \Delta_a ^ {(NP)} ( \mu^2 , \mu^2_0 )
\end{eqnarray}
where, in $\Delta_a ^ {(P)} $ , the integral over $z$ ranges from 0 to $z_{\text{dyn}}$, while in $\Delta_a ^ {(NP)}$, it ranges from $z_{\text{dyn}}$ to $z_M$.
The parameter $q_0$ is the minimal resolved emitted transverse momentum. 

\begin{figure}[!htb]
\centerline{
\includegraphics[width=0.8\textwidth]{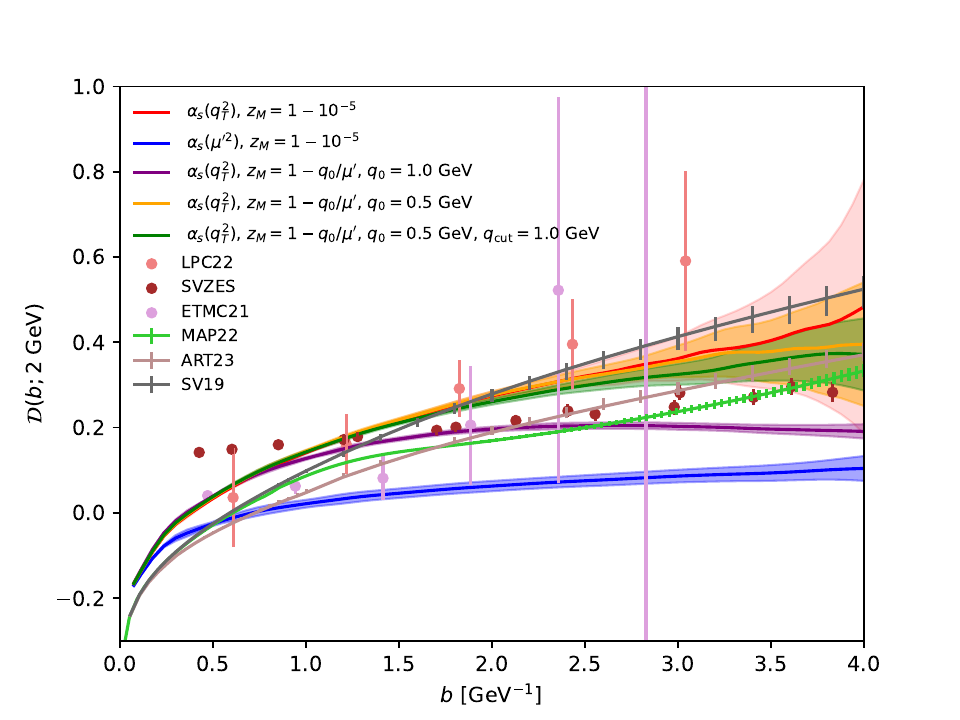}
}
\caption{CS kernels obtained from different PB models and several example extractions from the literature. (Plot taken from Ref. \cite{CSS1}).
}
\label{CS-kernel}
\end{figure}

Detailed introduction of all the variables are available in in Ref. \refcite{CSS1}. This separation allowed to illustrate an exact correspondence between the PB and Collins-Soper-Sterman (CSS) Sudakov form factors (available in different notation)\cite{Collins:1984kg,Collins:2011zzd}, for both perturbative and non-perturbative components. The accuracy of the PB Sudakov form factor was increased up to NNLL by including $A_a^{(3)}$ coefficient via the physical soft gluon coupling. It was observed that the non-perturbative part of the PB Sudakov form factor corresponds to the non-perturbative part of the CS kernel.

As shown in Fig.~\ref{CS-kernel}, the CS kernels were extracted from the PB approach for five different NNLL models, which differ in the amount of radiation controlled by the scale in $\alpha_s$ and the value of $z_M$. It was demonstrated that variations in the radiation models can lead to significantly different shapes of the extracted Collins-Soper (CS) kernel.

\section{Proper treatment of soft emissions by $z_{M}$ parameter }

In this section, I briefly review our latest studies on the importance of the proper treatment of soft emissions, as published in Refs.~\refcite{Mendizabal:2023mel,Bubanja:2023nrd,Bubanja:2024puv,Bubanja:2024crf}.
Ref.~\refcite{Mendizabal:2023mel} demonstrates that non-perturbative Sudakov form factors play a crucial role in inclusive distributions, such as collinear parton densities and Drell-Yan transverse momentum spectra. These soft emissions are essential to the $\overline{\text{MS}}$-scheme, as neglecting them would lead to the non-cancellation of important singular terms. 
It was also shown that these soft emissions have negligible impact on final-state hadron spectra and jets.

Ref.~\refcite{Bubanja:2023nrd} focuses on the determination of intrinsic-$k_T$, which is introduced into parton evolution as a non-perturbative parameter. For simplicity, the intrinsic-$k_T$ is modeled as a Gaussian distribution with a width $\sigma$, expressed as $e^{-k_T^2/\sigma^2}$, and multiplied by the parton density function at the starting scale. The study emphasizes the transverse momentum spectrum, $p_T(ll)$, and determines the width of the intrinsic-$k_T$ distribution through precise measurements at LHC energies \cite{CMS:2022ubq} across a wide range of Drell-Yan (DY) masses, $m_{DY}$. Lower-energy measurements were also analyzed, showing that the PB approach effectively describes all available data.

In the recent paper of the CMS Collaboration~\cite{CMS:2024goo}, the $\sqrt{s}$-dependence  of the intrinsic-$k_T$  width was confirmed for the event generators \textsc{PYTHIA} (for tunes CP3, CP4 and CP5 and \textsc{HERWIG} (for tunes CH1 and CH2). 

The predicton from \textsc{PYTHIA} and \textsc{HERWIG} event generator is in contrast to the \textsc{CASCADE3}, where only mild $\sqrt{s}$-dependence of the intrinsic-$k_T$ width is observed \cite{Bubanja:2023nrd}. 
We did two independent checks to understand from where thses differences are coming. 

\begin{figure}[!htb]
\centerline{
\includegraphics[width=0.7\textwidth]{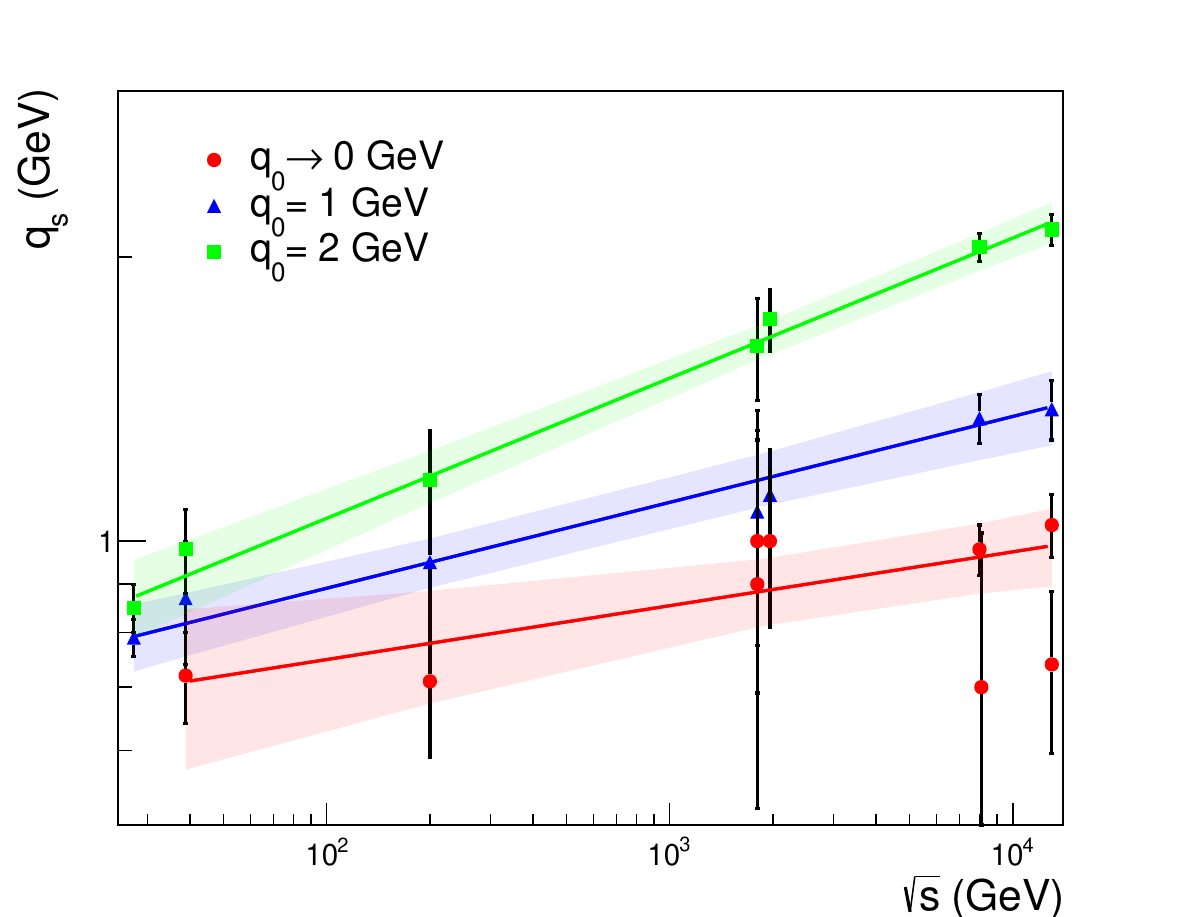}
}
\caption{The width parameter $q_s$ of the intrinsic-$k_T$ distribution as a function of $\sqrt{s}$ (Plot taken from Ref.~\cite{Bubanja:2024puv}).
}
 \label{kt-width}
 \vspace{0.5cm}
\end{figure}

\paragraph{\textsc{CASCADE3} study} In Ref.\refcite{Bubanja:2024puv}, we varied $q_0$ in our TMDs to produce predictions for different $q_0$ values. As shown in Fig.~\ref{kt-width}, excluding the non-perturbative Sudakov form factor by setting $q_0$ to 1 GeV or 2 GeV reveals a strong dependence of the width parameter $q_s$ on the center-of-mass energy $\sqrt{s}$. A larger $q_0$, which corresponds to a smaller $z_M$, results in a stronger dependence of $q_s$ on $\sqrt{s}$. This implies that at high $\sqrt{s}$, the lack of proper treatment of soft emissions is compensated by a larger, unphysical value of $q_s$. 
This study \cite{Bubanja:2024puv} confirms that the stable result reported in Ref.\refcite{Bubanja:2023nrd} is obtained using the original PB set, where $q_0 < 0.01$ GeV.

\paragraph{\textsc{PYTHIA} study} In Ref.~\refcite{Bubanja:2024crf}, we varied the ISR cut-off parameter within the range $0.5 < {\tt pT0Ref} < 2.0$~GeV in \textsc{PYTHIA}. The initial-state-radiation (ISR) cut-off scale, {\tt pT0Ref}, controls soft gluon emissions in \textsc{PYTHIA}.

\begin{figure} [htb!]
  \centering
\includegraphics[width=0.65\linewidth]{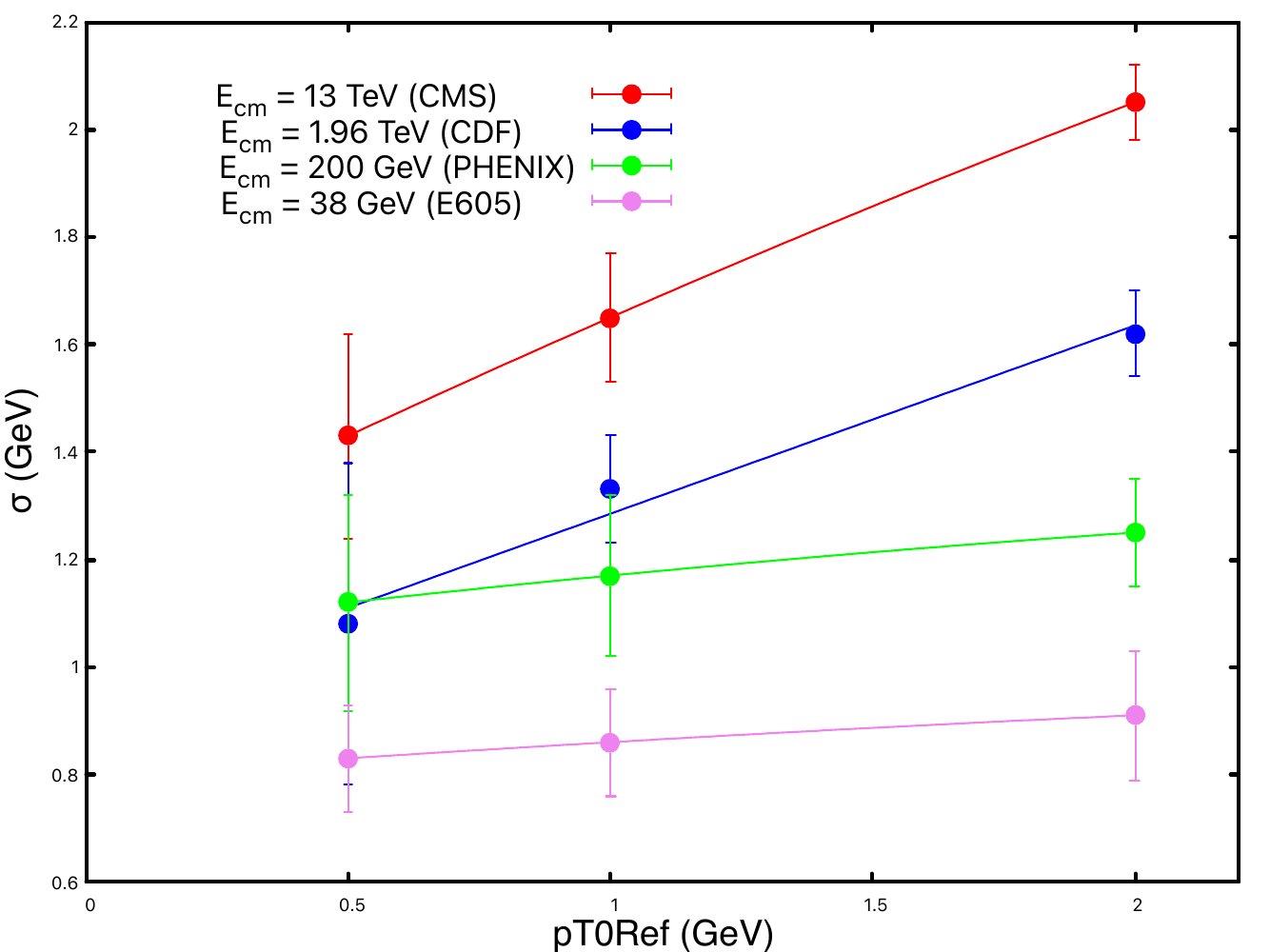} 
\caption {Dependence of the intrinsic-$k_T$ width, $\sigma$, on the ISR cut-off parameter, pT0Ref (Plot taken from Ref.~\cite{Bubanja:2024crf}).}  
\label{fig:qsvspt0ref}
\end{figure}

As shown in Fig.\ref{fig:qsvspt0ref}, we observed that the intrinsic $k_T$ width increases approximately linearly with the ISR cut-off parameter. This observation provides clear evidence that the $\sqrt{s}$-dependence is related to the no-emission probability, represented by the Sudakov form factor, through its dependence on the scale $z_{\text{dyn}}$, which is itself influenced by the ISR cut-off parameter. Furthermore, the $\mu$-dependent part of the Sudakov form factor was confirmed by examining the dependence of the width $\sigma$ on $m_{DY}$, which is directly linked to the evolution scale $\mu$ and becomes particularly evident at low $m_{DY}$.

These two parallel studies \cite{Bubanja:2024puv,Bubanja:2024crf} complete the puzzle of the intrinsic-$k_T$ dependence on $\sqrt{s}$. The next step is to improve \textsc{PYTHIA} based on the fundamental principles of the PB method.

\section{\textsc{PYTHIA} modification}

In this section, we present a summary of the results from Ref.~\refcite{PDF2ISR}. In that work, we introduced a method for constructing an initial-state parton shower model in which the backward evolution is fully consistent with the forward evolution of the collinear parton density. The method emphasizes the proper treatment of soft emissions in backward evolution, ensuring its alignment with the principles of forward evolution.

We refer to this approach as the {\scshape{Pdf2Isr}} method, which can be readily applied to any collinear parton density, provided the exact evolution conditions are specified. The {\scshape{Pdf2Isr}} method produces an initial-state parton shower that, in principle, is free of adjustable parameters and fully consistent with collinear parton densities at both LO and NLO.

\section*{Acknowledgments}
The works presented here were carried out in collaboration with I.~Bubanja, H.~Jung, A.~Lelek, L.~Lönnblad, M.~Mendizabal, N.~Raicevic, and K.~M.~Figueroa. I would like to thank them for their valuable contributions.

\end{document}